\documentstyle[emulateapj,psfig]{article}

\received{17 July 2000}  
\accepted{}

\def\etal{{\it et al.}}
\def\lsim{\hbox{ \rlap{\raise 0.425ex\hbox{$<$}}\lower 0.65ex\hbox{$\sim$} }}
\def\gsim{\hbox{ \rlap{\raise 0.425ex\hbox{$>$}}\lower 0.65ex\hbox{$\sim$} }}

\def\arcsec{\hbox{$^{\prime\prime}$}}

\def\f(h{\hbox{$~\!\!^{\rm h}$}}

\def\ale{\mathrel{\hbox{\rlap{\hbox{\lower4pt\hbox{$\sim$}}}\hbox{$<$}}}}
\def\age{\mathrel{\hbox{\rlap{\hbox{\lower4pt\hbox{$\sim$}}}\hbox{$>$}}}}

\slugcomment{Submitted to The Astrophysical Journal}

\lefthead{Bloom, Djorgovski, \& Kulkarni} 
\righthead{The redshift and the ordinary host galaxy of GRB 970228}

\begin{document}

\title{The Redshift and the ordinary host galaxy of GRB 970228\footnotemark}

\footnotetext{Partially based on the observations obtained at the
W.~M.~Keck Observatory which is operated by the California Association
for Research in Astronomy, a scientific partnership among California
Institute of Technology, the University of California and the National
Aeronautics and Space Administration.}

\author{J. S. Bloom, S. G. Djorgovski, S. R. Kulkarni}

\affil{Palomar Observatory 105--24, California Institute of Technology,
            Pasadena, CA 91125, USA; {\tt jsb,george,srk@astro.caltech.edu}}

\begin{abstract}
The gamma-ray burst of 28 February 1997 (GRB 970228) ushered in the
discovery of the afterglow phenomenon.  Despite intense study of the
host galaxy, however, the nature of the host and the distance to the
burst eluded the community.  Here we present the measurement of the
redshift of GRB 970228, and, based on its spectroscopic and
photometric properties, identify the host as an underluminous, but
otherwise normal galaxy at redshift $z=0.695$ undergoing a modest
level of star formation.  At this redshift, the GRB released an
isotropic equivalent energy of \hbox{$E = (6.8 \pm 0.5) \times
10^{51}\, h_{65}^{-2}$ erg} \hbox{[30--2000 keV restframe]}.  We find
no evidence that the host is much bluer or forming stars more
rigorously than the general field population. In fact, by all accounts
in our analysis (color--magnitude, magnitude--radius, star-formation
rate, Balmer-break amplitude) the host properties appear typical for
faint blue field galaxies at comparable redshifts.
\end{abstract}


\section{Introduction}

The gamma-ray burst of 28 February 1997 (hereafter GRB 970228) was a
watershed event, especially at optical wavelengths. The afterglow
phenomenon, long-lived multiwavelength emission, was discovered
following GRB 970228 in the X-rays (Costa \etal\ 1997\nocite{cfh+97})
and optical (\cite{vgg+97}).  Despite intense observations, no radio
transient of GRB 970228 was found (\cite{fksw98}); the first radio
afterglow (Frail \etal\ 1997\nocite{fkn+97}) had to await the next
BeppoSAX localization of GRB 970508. The basic predictions of the
synchrotron shock model for GRB afterglow appeared confirmed by GRB
970228 ({\it e.g.},~\cite{wrm97}).

Despite intense efforts, early spectroscopy of the afterglow GRB
970228 ({\it e.g.},~\cite{thcm97,kdc+97}) failed to reveal the
redshift.  Spectroscopy of the afterglow of GRB 970508 proved more
successful (Metzger \etal\ 1997), revealing that GRB originated from a
redshift $z \age 0.835$.  Through the preponderance of redshift
determinations and the association of GRBs with faint galaxies, it is
now widely believed that the majority or all of long duration ($T \age
1$ s) gamma-ray bursts originate from cosmological distances.

Even without a redshift, observations of the afterglow of GRB 970228 in
relation to its immediate environment began to shed light on the
nature of the progenitors of gamma-ray bursts.  Groundbased
observations of the afterglow revealed a near coincidence of the GRB
with the optical light of a faint galaxy (\cite{mkd+97,vgg+97}).
Later, deeper groundbased images (\cite{dkg+97}) and high resolution
images from the {\it Hubble Space Telescope} (HST) showed the light
from the fading transient clearly embedded in a faint galaxy
(\cite{slp+97}), the putative host of the GRB.  Further, these HST
images showed a measurable offset between the host galaxy centroid and
the afterglow.  Though by no means definitive, the offset of GRB
970228 rendered an active galactic nucleus (AGN) origin unlikely
(\cite{slp+97}).

The two most popular progenitor scenarios---coalescence of binary
compact stellar remnants and the explosion of a massive star
(``collapsar'')---imply that the gamma-ray burst rate should closely
follow the massive star formation rate in the Universe.  In both
formation scenarios a black hole is created as a byproduct; however,
the scenarios differ in two important respects. First, only very
massive progenitors ($M_{\rm ZAMS} \age 40\, M_\odot$; Fryer, Woosley,
\& Hartmann 1999\nocite{fwh99}) will produce GRBs in the single star
model whereas the progenitors of neutron star--neutron star binaries
need only originate with $M_{\rm ZAMS} \age 8\, M_\odot$.  Second, the
scenarios predict a distinct distribution of physical offsets ({\it
e.g.},~\cite{pac98,bsp99}) in that the coalescence site of merging
remnants could occur far from the binary birthplace (owning to
substantial systemic velocities acquired during neutron star formation
through supernovae) whereas exploding massive stars will naturally
occur in star-forming regions.  In relation to the predicted offset of
GRB 970228, Bloom \etal\ (1999)\nocite{bsp99} further noted the
importance of redshift to determine the luminosity (and infer mass) of
the host galaxy: massive galaxies more readily retain binary remnant
progenitors.  Thus the relationship of GRBs to their hosts is most
effectively exploited with redshift by setting the physical scale of
any observed angular offset and critically constraining the mass (as
proxied by host luminosity).

The redshift of GRB 970228 also plays a critical role in the emerging
supernova--GRB link.  Following the report of an apparent supernova
component in the afterglow of GRB 980326 (Bloom et
al.~1999\nocite{bkd+99}), the afterglow light curves of GRB 970228
were reanalyzed: both Reichart (1999\nocite{rei99}) and Galama et
al.~(2000\nocite{gtv+00}) found evidence for a supernova component.
This interpretation, however, relies critically on the knowledge of
the redshift to GRB 970228 to set the restframe wavelength of the
apparent broadband break of the SN component.

Finally, knowledge of redshift is essential to derive the physical
parameters of the GRB itself, primarily the energy scale.  We now
know, for instance, that the typical GRB releases about $10^{52}$ erg
in gamma-rays. The distribution of observed isotropic-equivalent GRB
energies is, however, very broad ({\it cf}.~Kulkarni et
al.~2000\nocite{kbb+00}).

Recognizing these needs we implemented an aggressive spectroscopy
campaign on the host of GRB 970228 as detailed in \S \ref{sec:obs}.
The redshift determination was first reported by Djorgovski et
al.~(1999) and is described in more detail in \S
\ref{sec:0228redshift}. We then use this redshift and the spectrum of
the host galaxy in \S \ref{sec:0228imp} to set the physical scale of
the observables: energetics, star-formation rates, and offsets.  Based
on this and photometric imaging from HST we demonstrate in \S
\ref{sec:host} that the host is an underluminous, but otherwise normal
galaxy.  Lastly, we compare the host with that of the field galaxy
population.

\section{Observations and Reductions}
\label{sec:obs}

Spectra of the host galaxy were obtained on the W.~M.~Keck Observatory
10 m telescope (Keck II) atop Mauna Kea, Hawaii. Observations were
conducted over the course of several observing runs: UT 1997 August
13, UT 1997 September 14, UT 1997 November 1 and 28--30, and UT 1998
February 21--24. The observing conditions were variable, from marginal
(patchy/thin cirrus or mediocre seeing) to excellent, and on some
nights no significant detection of the host was made; such data were
excluded from the subsequent analysis.  On most nights, multiple
exposures (2 to 5) of 1800 sec were obtained, with the object dithered
on the spectrograph slit by several arcsec between the exposures.  The
net total useful on-target exposure was approximately 11 hours from
all of the runs combined.

All data were obtained using the {\it Low-Resolution Imaging
Spectrometer} (LRIS; \cite{occ+95}) with 300 lines mm$^{-1}$ grating
and a 1.0 arcsec wide long slit, giving an effective instrumental
resolution FWHM $\approx 12$ \AA.  Slit position angle was always set
to $87^\circ$, with star S1 (\cite{vgg+97}) always placed on the slit,
and used to determine the spectrum trace along the chip; galaxy
spectra were then extracted at a position 2.8 arcsec east of the star
S1.  Efforts were made to observe the target at hour angles so as to
make this slit position angle as close to parallactic as possible.
Exposures of an internal flat-field lamp and arc lamps were obtained
at comparable telescope pointings immediately following the target
observations.  Exposures of standard stars from Oke \& Gunn
(1983)\nocite{og83} and Massey \etal\ (1998)\nocite{msb88} were
obtained and used to measure the instrument response curve, although
on some nights the flux zero points were unreliable due to
non-photometric conditions.

Wavelength solutions were obtained from arc lamps in the standard
manner, and then a second-order correction was determined from the
wavelengths of isolated strong night sky lines, and applied to the
wavelength solutions.  This procedure largely eliminates systematic
errors due to the instrument flexure, and is necessary in order to
combine the data obtained during separate nights.  The final
wavelength calibrations have the r.m.s.~$\sim 0.2 - 0.5$ \AA, as
determined from the scatter of the night sky line centers.  All
spectra were then rebinned to a common wavelength scale with a
sampling of 2.5 \AA\ (the original pixel scale is $\sim 2.45$ \AA),
using a Gaussian with a $\sigma = 2.5$ \AA\ as the
interpolating/weighting function.  This is effectively a very
conservative smoothing of the spectrum, since the actual instrumental
resolution corresponds to $\sigma \approx 5$ \AA.

Individual spectra were extracted and combined using a statistical
weighting based on the signal-to-noise ratio determined from the data
themselves (rather than by the exposure time).  Since some of the
spectra were obtained in non-photometric conditions, the final
spectrum flux zero-point calibration is also unreliable, but the
spectrum shape should be unaffected.  We use direct photometry of the
galaxy to correct this zero-point error (see below).

Our uncorrected spectrum gives a spectroscopic magnitude $V \approx
26.3$ mag for the galaxy.  Direct photometry from the HST data
indicates $V = 25.75 \pm 0.3$ (\cite{gtv+00}). Given that some of our
spectra were obtained through thin cirrus, this discrepancy is not
surprising.  Thus, in order to bring our measurements to a consistent
system, we multiply our flux values by a constant factor of 1.66, but
we thus also inherit the systematic zero-point error of $\sim 30$\%
from the HST photometry.

There is some uncertainty regarding the value of the foreground
extinction in this direction (see discussion in \S
\ref{sec:0228host}).  We apply a Galactic extinction correction by
assuming $E_{B-V} = 0.234$ mag from Schlegel, Finkbeiner \& Davis
(1998)\nocite{sfd98}. We assume $R_V = A_V/E_{B-V} = 3.2$, and the
Galactic extinction curve from Cardelli, Clayton \& Mathis
(1988)\nocite{ccm98} to correct the spectrum.  All fluxes and
luminosities quoted below incorporate both the flux zero-point and the
Galactic extinction corrections.

\section{The Redshift of GRB 970228}
\label{sec:0228redshift}

\begin{figure*}[tbp]
\centerline{\psfig{file=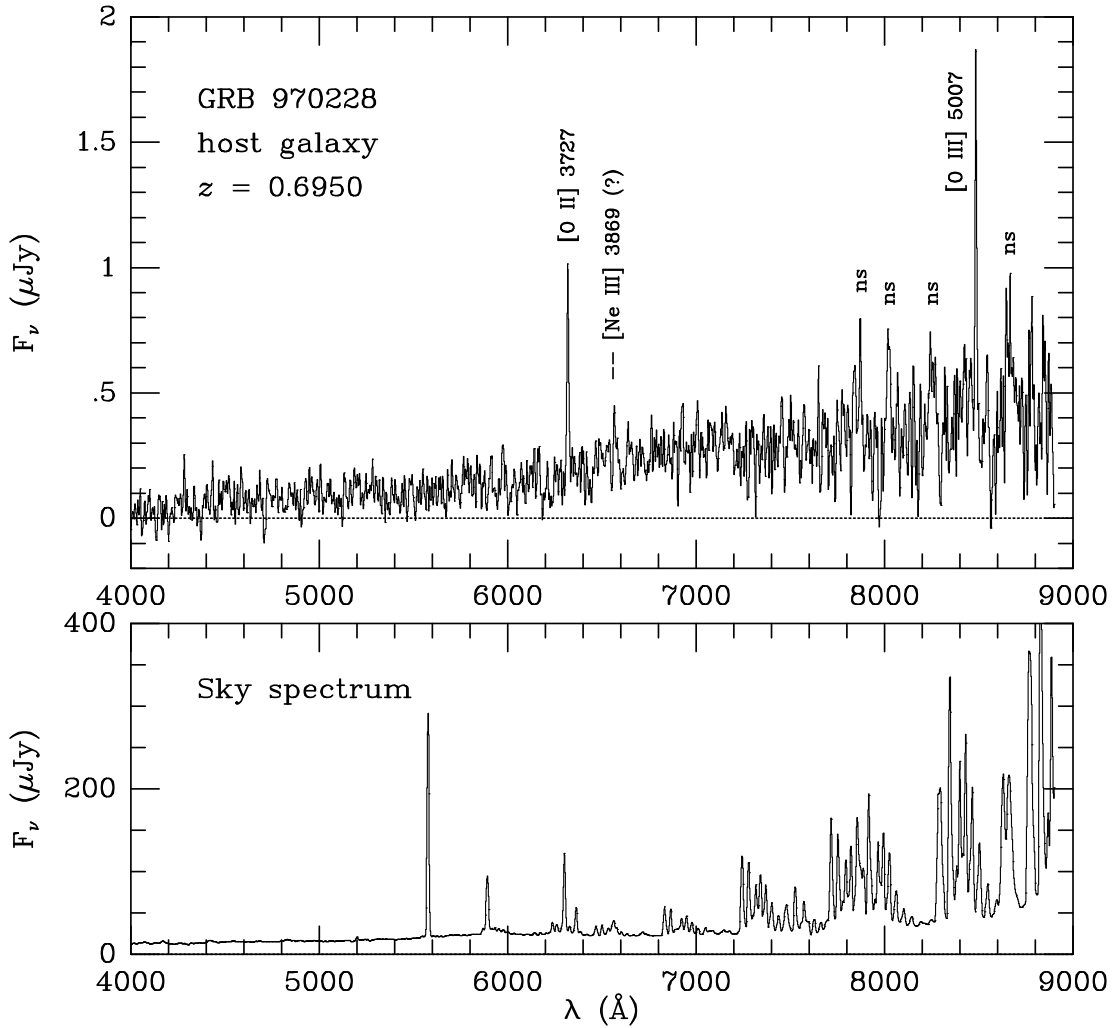,width=6.2in}}
\caption[]{(top) The weighted average spectrum of the host galaxy of
GRB 970228, obtained at the Keck II telescope.  Prominent emission
lines [O II] 3727 and [O III] 5007 and possibly [Ne III] 3869 are
labeled assuming the lines originate from the host at redshift $z =
0.695$. The notation ``ns'' refers to noise spikes from strong night
sky lines. (bottom) The average night sky spectrum observed during the
GRB 970228 host observations, extracted and averaged in exactly the
same way as the host galaxy spectrum. }
\label{fig:0228spec}
\end{figure*}

The final combined spectrum of the galaxy is shown in Figure
\ref{fig:0228spec}.  Two strong emission lines are seen, [O II] 3727
and [O III] 5007, thus confirming the initial redshift interpretation
based on the [O II] 3727 line alone (\cite{dkbf99a}).  Unfortunately
the instrumental resolution was too coarse to resolve the [O II] 3727
doublet. The weighted mean redshift is $z = 0.6950 \pm 0.0003$.  A
possible weak emission line of [Ne III] 3869 is also seen.
Unfortunately, the strong night sky OH lines preclude the measurements
of the H$\beta$ 4861 and [O III] 4959 lines, as well as the higher
Balmer lines.

The corrected [O II] 3727 line flux is $(2.2 \pm 0.1) \times 10^{-17}$
erg cm$^{-2}$ s$^{-1}$ Hz$^{-1}$, and its observed equivalent width is
$W_\lambda = 51 \pm 4$ \AA, {\it i.e.}, $30 \pm 2.4$ \AA\ in the
restframe.  This is not unusual for field galaxies in this redshift
range (\cite{hcbp98}).  The [Ne III] 3869 line, if real, has a flux of
at most 10\% of the [O II] 3727 line, which is reasonable for an
actively star forming galaxy. The corrected [O III] 5007 line flux is
$(1.55 \pm 0.12) \times 10^{-17}$ erg cm$^{-2}$ s$^{-1}$ Hz$^{-1}$,
and its observed equivalent width is $W_\lambda = 30 \pm 2$ \AA, {\it
i.e.}, $17.7 \pm 1.2$ \AA\ in the restframe. For the H$\beta$ line, we
derive an upper limit of $< 3.4 \times 10^{-18}$ erg cm$^{-2}$
s$^{-1}$ Hz$^{-1}$ ($\sim 1~\sigma$), and for its observed equivalent
width $W_\lambda < 7$ \AA.  We note however that this measurement may
be severely affected by the poor night sky subtraction.
 
The continuum flux at $\lambda_{obs} = 4746\,$\AA, corresponding to
$\lambda_{rest} = 2800\,$\AA, is $F_\nu = 0.29 ~\mu$Jy, with our
measurement uncertain by $\sim 10$\%.  The continuum flux at
$\lambda_{obs} \sim 7525\,$\AA, corresponding to the restframe $B$
band, is $F_\nu = 0.77 ~\mu$Jy, with our measurement uncertain by
$\sim 7$\%. Above, we report only the statistical uncertainties of all
fluxes; an additional systematic uncertainty of $\sim 30$\% is
inherited from the overall flux zero-point uncertainty.

\section{Implications of the redshift}
\label{sec:0228imp}

For the following discussion, we will assume a flat cosmology as
suggested by recent results ({\it e.g.},~\cite{dab+00}) with $H_0 =
65$ km s$^{-1}$ Mpc$^{-1}$, $\Omega_M = 0.3$, and $\Lambda_0 = 0.7$.
For $z = 0.695$, the luminosity distance is $1.40 \times 10^{28}$ cm,
and 1 arcsec corresponds to 7.65 proper kpc or 13.0 comoving kpc in
projection.

\subsection{Burst Energetics}

The gamma-ray fluence (integrated flux over time) is converted from
count rates under the assumption of a GRB spectrum, the spectral
evolution, and the true duration of the GRB.  These quantities are
estimated from the GRB data itself but can lead to large uncertainties
(a factor of few) in the fluence determination.  In table
\ref{tab:energetics} we summarize the fluence of GRB 970228 as
observed by all high energy experiments that detected the GRB. We
determine the implied energy release (col.~5, table
\ref{tab:energetics}) assuming isotropic emission. Further we
``standardize'' the energetics to the restframe 30--2000 keV in the
following manner.  We first normalize the observed fluences in a given
bandpass (col.~2 and 3) to a bandpass defined by $30 / (1 + z)$ to
$2000 / (1 + z)$ keV by a ratio of the integrated spectral shape over
these two bandpasses.  The implied energy release is then found
assuming isotropic emission and using the luminosity distance measure
for the assumed cosmology. If no spectral fit is reported we find the
median energy implied by assuming the spectral shape is each of the
average 54 spectra from Band \etal\ (1993\nocite{bmf+93}).  The
reported errors reflect the uncertainty in the redshift measure,
fluence, and spectral shape.

Given the significantly higher gamma-ray spectral and timing
resolution of the TGRS instrument relative to the others, we favor the
isotropic energy implied by the TGRS analysis: \hbox{$E = (6.8 \pm
0.5) \times 10^{51}\, h_{65}^{-2}$ erg} [30--2000 keV restframe].
That the implied energy is a factor of $\sim 2$---3 higher using
measurements from BeppoSAX and Ulysses reflects the importance of high
signal-to-noise spectroscopy in ascertaining the fluence and hence the
energy release. The slow decline and absence of a strong break in the
optical light curve ({\it e.g.},~Galama \etal\ 1997\nocite{ggv+97})
suggests that the GRB emission was nearly isotropic (see also
\cite{sph99}) and so the knowledge of $E$ is primarily limited by the
accuracy of the fluence measurement.

\subsection{The offset of the GRB and the host morphology}
\label{sec:off}

\begin{figure*}[tbp]
\centerline{\psfig{file=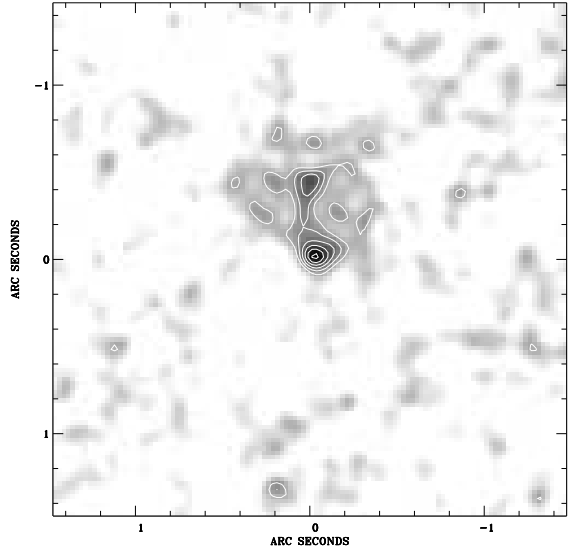,width=4.2in}}
\caption[]{A 3\arcsec $\times$ 3\arcsec\ (23 $\times$ 23 kpc$^2$ in
projection) region of the HST/STIS image (4.7 September 1997 UT) of
the host galaxy of GRB 970228.  The image has been smoothed (see text)
and is centered on the optical transient.  North is up and East is to
the left.  Contours in units of 3,4,5,6,7,8 background $\sigma$
($\sigma = 2.41$ DN) are overlaid.  The transient is found on the
outskirts of detectable emission from a faint, low-surface brightness
galaxy.  The morphology is clearly not that of a classical Hubble type,
though there appears to be a nucleus and an extended structure to the
north of the transient.}
\label{fig:host}
\end{figure*}

For the purpose determining the position of the GRB within its host,
we examined the HST/STIS observations taken on 4.7 Sept 1997 UT
(\cite{fpt+99}).  The observation consisted of eight 575$\,$s STIS
clear (CCD50) exposures paired in to four 1150s to facilitate removal
of cosmic rays.  We processed these images using the drizzle technique
of Fruchter \& Hook (1997)\nocite{fh97} to create a final image with a
plate scale of 0.0254 arcsec pixel$^{-1}$.  To enhance the low-surface
brightness host galaxy we smoothed this image with a Gaussian with
$\sigma = 0.043$ arcsec. The optical transient is well-detected in
figure \ref{fig:host} (point source towards the South) and clearly
offset from the bulk of the detectable emission of the host.

Two morphological features of the host stand out: a bright knot
manifested as an sharp 6-$\sigma$ peak near the centroid of the host
to north of the transient and an extension from this knot towards the
transient. Although, as we demonstrate below, this host is a
subluminous galaxy ({\it i.e.},~not a classic late-type $L_*$ spiral
galaxy) we attribute these features to a nucleus and a spiral-arm,
respectively. It is not unusual for dwarf galaxies to exhibit these
canonical Hubble-diagram structures (S.~Odewahn, private
communication).  These feature have not previously been noted in the
literature.

Centroiding the transient and the nucleus components within a 3 pixel
aperture radius about their respective peak, we find an angular offset
of $436 \pm 14$ milliarcsec between the nucleus and the optical
transient.  With our assumed cosmology, this amounts to a projected
physical distance of $3.34\, \pm\, 0.11\, h_{65}^{-1}$ kpc.

\subsection{Physical Parameters of the host galaxy}
\label{sec:0228host}

We found half-light radius of the host galaxy using our final drizzled
HST/STIS image: we mask a 3 $\times$ 3 pixel region around the
position of the optical transient and inspect the curve--of--growth
centered on the central bright knot, the supposed nucleus and estimate
the half-light radius to be 0.31 arcsec or \hbox{2.4 $h_{65}^{-1}$
kpc} (physical) at a redshift of $z=0.695$. The half-light radius
visually estimated from curve--of--growth in the WFPC F814W and F606W
filters (see \cite{cl99c}) is comparable.

Although there is some debate (at the 0.3 mag level in $A_V$) as to
the proper level of Galactic extinction toward GRB 970228
(\cite{cl99c,gfd99,fpt+99}) we have chosen to adopt the value $E(B-V)
= 0.234$ found from the dust maps of \cite{sfd98} and a Galactic
reddening curve $R_V = A_V/E(B-V) = 3.2$.  Using extensive reanalysis
of the {\it HST} imaging data by Galama \etal\ 2000\nocite{gtv+00} the
extinction corrected broadband colors of the host galaxy as $V = 25.0
\pm 0.2$, $R_c = 24.6 \pm 0.2$, $I_c = 24.2 \pm 0.2$. These measures,
consistent with those of Castander \& Lamb (1999)\nocite{cl99c} and
Fruchter \etal\ (1999)\nocite{fpt+99}, are derived from the WFPC2
colors and broadband STIS flux. The errors reflect both the
statistical error and the uncertainty in the spectral energy
distribution of the host galaxy.  We have not included a contribution
from the uncertainty in the Galactic extinction. Using the NICMOS
measurement from Fruchter \etal\ (1999)\nocite{fpt+99} the extinction
corrected infrared magnitude is $H_{\rm AB} = 24.6 \pm 0.1$.  Using
the zero-points from Fukugita \etal\ (1996)\nocite{fig+96}, the
extinction-corrected AB-magnitudes of the host galaxy are: $V_{\rm AB}
= 25.0$, $R_{\rm AB} = 24.8$, $I_{\rm AB} = 24.7$.

To facilitate comparison with moderate redshift galaxy surveys (\S
\ref{sec:host}) we compute the restframe $B$-band magnitude of the
host galaxy.  From the observed continuum in the restframe $B$ band,
we derive the absolute magnitude $M_B = -18.4 \pm 0.4$ mag [or
\hbox{$M_{\rm {AB}}(B) = -18.6 \pm 0.4$ mag}], {\it i.e.}, only
slightly brighter than the LMC now.  For our chosen value of $H_0$, an
$L_*$ galaxy at $z \sim 0$ has $M_B \approx -20.9$ mag, and thus the
host at the observed epoch has $L \sim 0.1~L_*$ today.  Its observed
morphology from the HST images is also consistent with a dwarf,
low-surface brightness galaxy.

\subsection{Star formation in the host}

From the [O II] 3727 line flux, we derive the line luminosity $L_{3727} =
5.44 \times 10^{40}$ erg s$^{-1}$ ($\pm 5$\% random) ($\pm 30$\%
systematic).  Using the star formation rate estimator from
\cite{ken98}, we derive the SFR $\approx 0.76 ~M_\odot$ yr$^{-1}$.
Using a 3-$\sigma$ limit on the H$\beta$ flux, we estimate $L_{H\beta}
< 2.5 \times 10^{40}$ erg s$^{-1}$.  Assuming the H$\alpha$/H$\beta$
ratio of $2.85 \pm 0.2$ for the Case B recombination and a range of
excitation temperatures, we can derive a pseudo-H$\alpha$ based
estimate of the star formation rate ({\it cf.}~\cite{ken98}), SFR $<
0.6~M_\odot$ yr$^{-1}$, but we consider this to be less reliable than
the [O II] 3727 measurement.  From the UV continuum luminosity at
$\lambda_{rest} = 2800$\AA, following Madau, Pozzetti \& Dickinson
(1998)\nocite{mpd98}, we derive SFR $\approx 0.54 ~M_\odot$ yr$^{-1}$.
 
We note that the net uncertainties for each of these independent SFR
estimates are at least 50\%, and the overall agreement is encouraging.
While we do not know the effective extinction corrections in the host
galaxy itself, these are likely to be modest, given its blue colors
({\it cf.}~\S \ref{sec:host}), and are unlikely to change our results by
more than a factor of two.  (We hasten to point out that we are
completely insensitive to any fully obscured star formation component,
if any is present.)  On the whole, the galaxy appears to have a rather
modest (unobscured) star formation rate, $\sim 0.5 - 1~M_\odot$
yr$^{-1}$.  Given the relatively normal equivalent width of the [O II] 3727
line, even the star formation per unit mass does not seem to be
extraordinarily high.
 
\section{The Nature of the Host Galaxy}
\label{sec:host}

At $M_{\rm AB}(B) = -18.6$, the host galaxy of GRB 970228 is a
sub-luminous galaxy roughly 2.7 mag below $L_*$ at comparable
redshifts (\cite{lth+95}).  Galama \etal\ (2000), based on the
redshift of GRB 970228, recently found that an Sc galaxy spectral
energy distribution reasonably fits the optical--IR photometric fluxes
of the host galaxy.  This differs from the analysis of Castander \&
Lamb (1999)\nocite{cl99c} which would, now given the redshift of $z =
0.695$, favor a classification of an ``Irregular'' galaxy having
undergone burst of star-formation over the past few hundred Myr.
Clearly it is difficult to precisely determine the galaxy type without
more precise photometry and knowledge of the true Galactic extinction,
but our identification of a nucleus and possible arm structure (\S
\ref{sec:off}) supports the idea that the host is a late-type
dwarf. Indeed, the host has similar characteristics to that of the
Large Magellanic Cloud.  We further note that compared with the Simard
\etal\ (1999)\nocite{skf+99}, the magnitude-size relation of the host
galaxy is consistent with that observed for late-type and dwarf-irregular
galaxies.

\begin{figure*}[tbp]
\centerline{\psfig{file=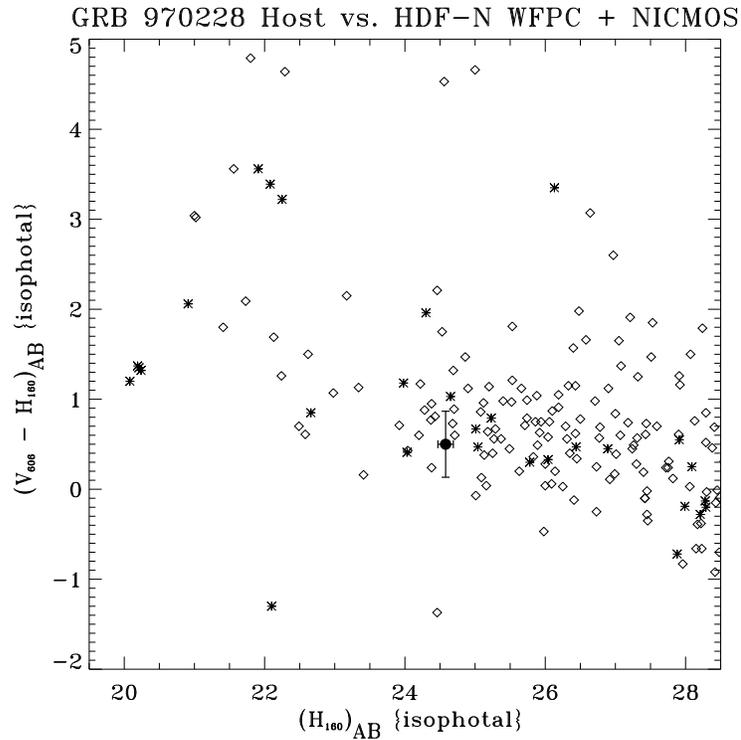,width=4.2in}}
\caption[]{Comparison of the color-magnitude of the host galaxy of GRB
970228 with the Hubble Deep Field North (HDF-N).  No systematic
difference, assuming a Galactic extinction towards GRB 970228 of $A_V
= 0.75$, is found between field galaxies at comparable magnitudes and
the host (denoted as a solid circle with error bars).  NICMOS and WFPC
photometry are taken from Thompson \etal\ (1999) and Williams \etal\
(1996)\nocite{wbd+96}, respectively.  The diamonds ($\diamond$)
represent extended objects (with the ratio of semi-major to semi-minor
axes less than 0.9) and the asterisks ($\ast$) compact galaxies and
stars. The error bars on the HDF-N data have been suppressed. See the
text for an explanation of the selection criteria.}
\label{fig:hdf}
\end{figure*}

The flat continuum suggests little restframe extinction in this
galaxy. We found the isophotal $(H - V)_{\rm AB}$ color of galaxies in
the Hubble Deep Field North (HDF-N) using the published photometry
from Thompson \etal\ (1999)\nocite{tsw+99} (NICMOS: F160W) and
Williams \etal\ (1996)\nocite{wbd+96} (WFPC4: F606W filter).  All WFPC
object identifications within 0.3 arcsec of a NICMOS identification
are plotted in figure \ref{fig:hdf} along with the dereddened color of
the host galaxy of GRB 970228.  Our field selection essentially biases
the color--magnitude relation towards redder objects and would serve
to accentuate the locus of the field with a blue galaxy.  Even with
this bias, there is no indication that the host is substantially more
blue than field galaxies at comparable magnitudes.

This conclusion---that the host galaxy of GRB 970228 is not
exceptionally blue---is at odds with that of Fruchter \etal\ (1999) who
have claimed that the host galaxy is unusually blue as compared with
typical field galaxies.  The difference may be due to the fact that
the Fruchter \etal\ (1999) analysis compared the host colors with a
significantly more shallow infrared survey than the NICMOS HDF
essentially masking the trend for faint galaxies to appear more blue.

\begin{figure*}[thp]
\centerline{\psfig{file=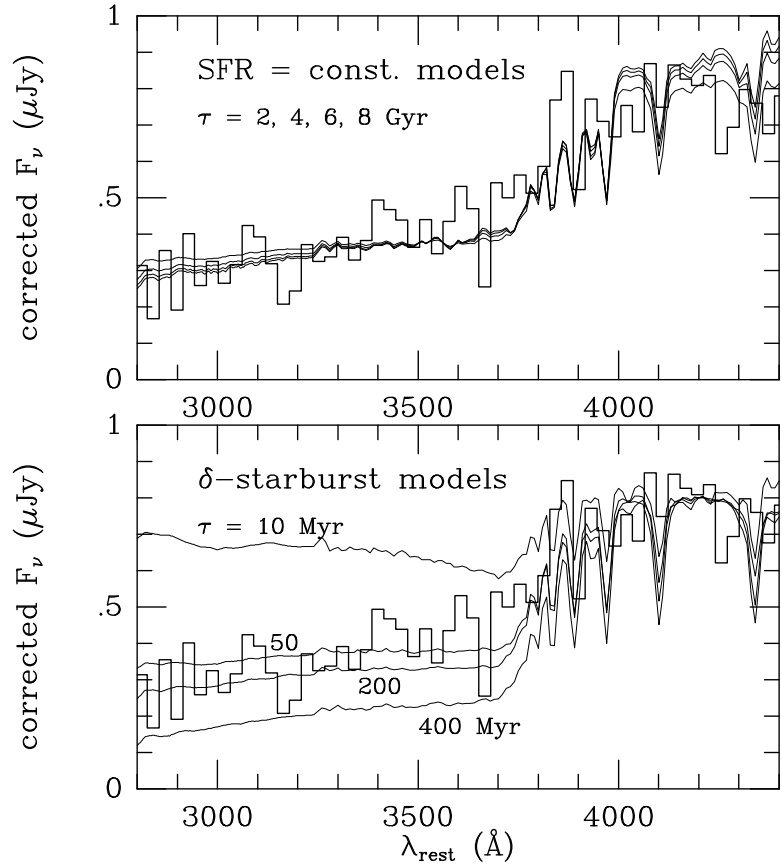,width=4.2in}}
\caption[]{Median-binned portion of the host spectrum near the Balmer
decrement. (top panel) Overlaid are Bruzual \& Charlot
(1993)\nocite{bc93} galaxy synthesis models assuming a varying time of
constant star formation. (bottom panel) Overlaid are \cite{bc93}
galaxy synthesis models assuming an instantaneous burst of
star-formation having occurred $\tau$ years since observation. Clearly
the host continuum could not be dominated by a young population of
stars ($\tau = 10$ Myr).  See the text for a discussion.  }
\label{fig:0228balmer}
\end{figure*}

Figure \ref{fig:0228balmer} shows a section of the median-binned
spectrum of the host galaxy.  The Balmer break is clearly detected,
with an amplitude $\Delta m \approx 0.55$ mag, which is typical for
the Balmer break selected population of field galaxies at $z \sim 1$
(K.~Adelberger, private communication).  For reference we also plot
several population synthesis model spectra (\cite{bc93}). The top
panel shows model spectra for a galaxy with a uniform star formation
rate, which may be a reasonable time-averaged approximation for a
normal late-type galaxy.  The correspondence is reasonably good and
does not depend on the model age. The bottom panel shows models with
an instantaneous burst of star formation.  In order to match the data,
we require fine-tuning of the post-burst age to be $\sim 10^{8} \times
2^{\pm 1}$ yr. No attempt was made to optimize the fit or to seek best
model parameters and the purpose of this comparison is simply
illustrative.  Clearly, if there was an ongoing or very recent burst
of star formation, the spectrum would be much flatter, with a weaker
Balmer break.
 
\section{Discussion and conclusion}

We have determined the redshift of the host galaxy of GRB 970228 to be
$z=0.695$ based on [O II] 3727 and [O III] 5007 line emission.  The
implied energy release $(6.8 \pm 0.5) \times 10^{51}$ [30---2000 keV
restframe] is on the smaller end of, but still comparable to, the
handful of other bursts with energy determinations ({\it
e.g.},~Kulkarni \etal\ 2000).  The absence of a detectable break in
the afterglow light curve we take to imply that any collimation of
emission ({\it i.e.},~jetting) neglegibly reduces the estimate of
total energy release in GRB 970228 (although \cite{fwk00}, using
late-time radio data, have found that even without an optical break,
GRB 970508 may have been collimated).

Most GRB transients appear spatially coincident with faint host
galaxies, disfavoring the merging NS hypothesis ({\it e.g.},~\cite{pac98}).
The coincidence of GRB 970508 with its host (\cite{fp98,bdkf98}) is
particularly constraining given the excellent spatial coincidence of
the GRB with the center of a dwarf galaxy ({\it cf.}~\cite{bdkf98}).  The
transient of GRB 970228 lies $3.34 \pm 0.11\, h_{65}^{-1}$ kpc from
the center of the galaxy, about 1 kpc in projection outside the
half-light radius of the galaxy.  From the above analysis we have
shown, like GRB 970508, the host is underluminous ($L \approx 0.05
L_*$), and by assumption, undermassive relative to $L_*$ galaxies.
According to Bloom \etal\ (1999)\nocite{bsp99}, about 50\% of merging
neutron binaries should occur beyond 3.5 kpc in projection of such
dwarf galaxies.  Thus, by itself, the offset of GRB 970228 from its
host does not particularly favor a progenitor model.

With a star-formation rate of \hbox{0.5--1 $M_\odot$ yr$^{-1}$}, the
host of GRB 970228 is forming stars only at a moderately higher rate
than comparable galaxies in the local universe.  This result is not
surprising given that, on a whole, the universal rate of
star-formation increases steeply at least out until redshift $z
\approx 1.5$.  If the association of GRBs with massive stars is
correct, however, GRB hosts should reveal an increased propensity to
form massive stars over and above their counterpart field
galaxies. Moreover, the massive star formation rate should be vigorous
at the time GRB occurs since massive stars require a negligible time
(compared to typical star-formation burst durations) to explode since
zero-age main sequence.

In $R$-band magnitude the host is near the median of GRB hosts
observed to date but in absolute $B$-band magnitude the host at the
faint end of the distribution. Of the host galaxies detected thus far
only GRB 970508 is as comparably faint to the host of GRB
970228. Except in angular extent, the host galaxies of GRB 970228 and
GRB 970508 (\cite{bdkf98}) bear a striking resemblance.  Both appear
to be sub-luminous ($L \ale 0.1 L_*$), compact and blue.  Spectroscopy
of both reveal the presence of the [Ne III] 3869 line, indicative of
recent very massive star formation. However, such properties are not
shared by all of the GRB hosts studied to date. We note too the rather
curious trend that the two GRBs themselves appear to have similar
properties in that they decay slowly, are the two least luminous in
term of GRB energetics, and do not exhibit evidence of a strong break
in the light curve.

\acknowledgments

The authors thank the generous support of the staff of the W.~M.~Keck
Foundation.  This paper has benefited from stimulating conversations
with P.~van Dokkum and K.~Adelberger. We thank M.~van Kerkwijk for
help during observing and C.~Clemens for his use of dark-time
observing nights.  JSB gratefully acknowledges the fellowship from the
Fannie and John Hertz Foundation. SGD acknowledges partial funding
from the Bressler Foundation. This work was supported in part by
grants from the NSF and NASA to SRK.


\supereject
\begin{table*}
\caption[]{Implied Energetics of GRB 970228\label{tab:energetics}}
\begin{tabular}{lccccr}
\hline
\hline
Instrument & Bandpass & $S_{-6}^a$  & $E_{30-2000}^b$  & Refs. \\
       &  [keV]   & [erg cm$^{-2}$]  & [$\times 10^{51}$ erg] & \\
\hline
TGRS/WIND\nocite{stc+96}
  & 50--300 & $3.1 \pm 0.2$ & 
	$6.8 \pm 0.5$ &       
  1,2 \nocite{stc+96,pcg+98} \\ 

GRBM+WFC\nocite{fcf+97,jmb+97} 
  & 40--700 & $11 \pm 1$  & $21 \pm 2$ 
& 3,4,5 \nocite{fcf+97,jmb+97,fcp+98} \\
~~~/BeppoSAX & 50--300 & $6.1$      
& $20 \pm 6$ 
& \\
 
\medskip
GRB/Ulysses\nocite{hsa+92} & 25--100 & $4.3$ & 31 $\pm$ 16 & 6, 7\nocite{hsa+92,hcf+97} \\
\hline
\end{tabular}
\raggedright

\noindent {\uppercase{REFERENCES}---} 1. \cite{stc+96} 2. \cite{pcg+98} 3.
\cite{fcf+97} 4. \cite{jmb+97} 5. \cite{fcp+98} 6. \cite{hsa+92}
7. \cite{hcf+97} 

\medskip 
$^a$ Fluence ($\times 10^6$) in given bandpass over
the duration of the GRB as determined from spectral fits.

$^b$ Implied energy release in GRB 970228 in the bandpass 30--2000 keV
restframe assuming isotropic emission. We have assumed a cosmology
with $H_0 = 65$ km s$^{-1}$ Mpc$^{-1}$, $\Omega_M = 0.3$, and
$\Lambda_0 = 0.7$. See text for explanation.

\end{table*}
\smallskip

\end{document}